\documentclass[pra,twocolumn,floatfix,superscriptaddress]{revtex4}

\usepackage{graphicx}
\usepackage{color}
\usepackage{subfigure}

\usepackage[normalem]{ulem}
\usepackage{braket}
\usepackage{amsmath}
\usepackage{verbatim}
\usepackage{amssymb}
\usepackage{mathtools}
\usepackage{bm}

\usepackage[pdfstartview=FitH]{hyperref}
\hypersetup{
    colorlinks=true,       
    linkcolor=cyan,          
    citecolor=magenta,        
    filecolor=magenta,      
    urlcolor=cyan,           
    runcolor=cyan
}

\newcommand{\beq}{\begin{equation}}
\newcommand{\eeq}{\end{equation}}
\newcommand{\bea}{\begin{eqnarray}}
\newcommand{\eea}{\end{eqnarray}}
\newcommand{\nn}{\nonumber \\}
\newcommand{\E}{\mathbb{E}}
\newcommand{\Ham}{\mathcal{H}}
\newcommand{\z}{\mathcal{\bzeta}}
\def\pd{\partial} 
\def\ph{\phantom{\hspace{1cm}}}
\def\x{\mathbf{x}}
\def\z{\mathbf{z}}
\def\bzeta{\bm{\zeta}}
\def\btheta{{\bm{\theta}}}
\def\bepsilon{{\bm{\rho}}}
\def\brho{{\bm{\rho}}}
\def\bphi{{\bm{\phi}}}

\def\tr{\text{Tr}}
\def\Ex{\mathbb{E}_{\x \sim p_\text{data}}}

\begin{document}

\title{Quantum Variational Autoencoder}

\author{Amir Khoshaman\footnote{Both authors contributed equally to this work.\label{note}}}
\affiliation{D-Wave Systems Inc., 3033 Beta Avenue, Burnaby BC
Canada V5G 4M9}

\author{Walter Vinci\textsuperscript{\ref*{note}}}
\affiliation
{D-Wave Systems Inc., 3033 Beta Avenue, Burnaby BC
Canada V5G 4M9}

\author{Brandon Denis}
\affiliation
{D-Wave Systems Inc., 3033 Beta Avenue, Burnaby BC
Canada V5G 4M9}

\author{Evgeny Andriyash}
\affiliation
{D-Wave Systems Inc., 3033 Beta Avenue, Burnaby BC
Canada V5G 4M9}

\author{Hossein Sadeghi}
\affiliation
{D-Wave Systems Inc., 3033 Beta Avenue, Burnaby BC
Canada V5G 4M9}

\author{Mohammad H. Amin }
\affiliation{D-Wave Systems Inc., 3033 Beta Avenue, Burnaby BC
Canada V5G 4M9}
\affiliation{Department of Physics, Simon Fraser
University, Burnaby, BC Canada V5A 1S6}

\begin{abstract}

Variational autoencoders (VAEs) are powerful generative models with the salient ability to perform inference. Here, we introduce a \emph{quantum variational autoencoder} (QVAE): a VAE whose latent generative process is implemented as a quantum Boltzmann machine (QBM). We show that our model can be trained end-to-end by maximizing a well-defined loss-function: a ``quantum" lower-bound to a variational approximation of the log-likelihood. We use quantum Monte Carlo (QMC) simulations to train and evaluate the performance of QVAEs. To achieve the best performance, we first create a VAE platform with discrete latent space generated by a restricted Boltzmann machine (RBM). Our model achieves state-of-the-art performance on the MNIST dataset when compared against similar approaches that only involve discrete variables in the generative process. 
We consider QVAEs with a smaller number of latent units to be able to perform QMC simulations, which are computationally expensive. We show that QVAEs can be trained effectively in regimes where quantum effects are relevant despite training via the quantum bound. Our findings open the way to the use of quantum computers to train QVAEs to achieve competitive performance for generative models. Placing a QBM in the latent space of a VAE leverages the full potential of current and next-generation quantum computers as sampling devices.

\end{abstract}

\maketitle
\section{Introduction}
\label{sec:intro}

While rooted in fundamental ideas that date back decades ago~\cite{rosenblatt1958perceptron,rumelhart1988learning}, deep-learning algorithms~\cite{hinton2006fast,bengio2007greedy} have only recently started to revolutionize the way information is collected, analyzed, and interpreted in almost every intellectual endeavor~\cite{lecun2015deep}. This is made possible by the computational power of modern dedicated processing units (such as GPUs). The most remarkable progress has been made in the field of supervised learning~\cite{goodfellow2016deep}, which requires a labeled dataset.   There has also been a  surge of interest in the development of unsupervised learning with unlabeled data~\cite{vincent2008extracting,hinton1995wake,bengio2007greedy}. One notable challenge in unsupervised learning is the computational complexity of training most models~\cite{tieleman2008training}. 

It is reasonable to hope that some of the computational tasks required to perform both supervised and unsupervised learning could be significantly accelerated by the use of \emph{quantum processing units} (QPU). Indeed, there are already quantum algorithms that can accelerate machine learning tasks~\cite{harrow2009quantum,wiebe2012quantum,childs2015quantum,lloyd2014quantum}. Interestingly, machine learning algorithms have been used in quantum-control techniques to improve fidelity and coherence~\cite{dolde2013high,zahedinejad2016designing,las2016genetic}. This natural interplay between machine learning and quantum computation is stimulating a rapid growth of a new research field known as  \emph{quantum machine learning}~\cite{schuld2015introduction,wittek2014quantum,adcock2015advances,arunachalam2017survey,Biamonte:2017db}.

A full implementation of quantum machine-learning algorithms requires the construction of fault-tolerant QPUs, which is still challenging~\cite{nielsen2002quantum,lidar2013quantum,fowler2012surface}. However, the remarkable recent development of gate-model processors with a few dozen qubits~\cite{neill2017blueprint,kandala2017hardware} and quantum annealers with a few thousand qubits~\cite{farhi2001quantum,johnson2011quantum} has triggered an interest in developing quantum machine-learning algorithms that can be practically tested on current and near-future quantum devices. Early attempts to use small gate-model devices for machine learning use techniques similar to those developed in the context of quantum approximate optimization algorithms (QAOA)~\cite{farhi2014quantum} and variational quantum algorithms (VQA)~\cite{peruzzo2014variational,kandala2017hardware} to perform quantum heuristic optimization as a subroutine for small unsupervised tasks such as clustering~\cite{otterbach2017unsupervised}. The use of quantum annealing devices for machine-learning tasks is perhaps more established and relies on the ability of quantum annealers to perform both optimization~\cite{kadowaki1998quantum,santoro2002theory,brooke2001tunable} and sampling~\cite{amin2015searching,venuti2016adiabaticity,albash2017temperature,vinci2017scalable}.

As optimizers, quantum annealers have been used to perform supervised tasks such as classification~\cite{neven2008training,denchev2012robust,pudenz2013quantum,mott2017solving}.  As samplers, they have been used to  train  RBMs, and are thus well-suited to perform unsupervised tasks such as training deep probabilistic models~\cite{denil2011toward,raymond2016global,korenkevych2016benchmarking,perdomo2017opportunities}.  In Ref.~\cite{adachi2015application}, a D-Wave quantum annealer was used to train a deep network of stacked RBMs to classify a coarse-grained version of the MNIST dataset~\cite{lecun1998mnist}. Quantum annealers have also been used to train fully visible Boltzmann machines  on small synthetic datasets~\cite{benedetti2016estimation,benedetti2016quantum}. While mostly used in conjunction with traditional RBMs, quantum annealing should find a more natural application in the training of QBM~\cite{Amin:2016qd}. 

A clear disadvantage of such early approaches is the need to consider datasets with a  small number of input units, which prevents a clear route towards practical applications of quantum annealing with current and next-generation devices. A first attempt towards this end was presented in Ref.~\cite{benedetti2018quantum}, with the introduction of a quantum-assisted Helmholtz machine (QAHM). However,  training QAHM is based on the \emph{wake-sleep} algorithm~\cite{hinton1995wake}, which does not have a well-defined loss function in the wake and sleep phases of training. Moreover, the gradients do not correctly propagate in the networks between the two phases. Because of these shortcomings, QAHM generates blurry images and training does not scale to standard machine-learning datasets such as MNIST.

Our approach is to use variational auto-encoders (VAEs), a class of generative models that provide an efficient inference mechanism~\cite{kingma2013auto,burda2015importance}. We show how to implement a  quantum VAE (QVAE), \emph{i.e.}, a VAE with discrete variables (DVAE)~\cite{rolfe2016discrete} whose generative process is realized by a QBM. QBMs were introduced in Ref.~\cite{Amin:2016qd}, and can be trained by minimizing a quantum lower  bound to the true log-likelihood. We show that QVAEs can be effectively trained by sampling from the QBM with continuous-time quantum Monte Carlo (CT-QMC).  We demonstrate that QVAEs have performance on par with  conventional DVAEs equipped with traditional RBMs, despite being trained via an additional bound to the likelihood.

QVAEs share some similarities with QAHMs, such as the presence of both an \emph{inference (encoder)} and a \emph{generation (decoder)} network. However, they have the advantage of a well-defined loss function with fully propagating gradients that can be efficiently trained via backpropagation. This allows to  achieve state-of-the-art performance (for models with only discrete units) on standard datasets such as MNIST by training (classical) DVAEs with large RBMs. Training QVAEs with a large number of latent units is  impractical with CT-QMC, but can be be accelerated with quantum annealers. Our work thus opens a path to practical machine learning applications with current and next-generation quantum annealers.

The QVAEs we introduce in this work are generative models with a classical autoencoding structure and a quantum generative process. This is in contrast with the quantum autoencoders (QAEs) introduced in Refs.~\cite{romero2017quantum,wan2017quantum}. QAEs have a quantum autoencoding structure (realized via quantum circuits), and can be used for quantum and classical data compression, but lack a generative structure.

The structure of the paper is as follows. In Sec.~\ref{sec:GEN} we provide a general discussion of generative models with latent variables, which include VAEs as a special case. We then introduce the basics of VAEs with continuous latent variables in Sec.~\ref{sec:VAE}. Sec.~\ref{sec:DVAE} discusses the generalization of VAEs to discrete latent variables and presents our experimental results with RBMs implemented in the latent space. In Sec.~\ref{sec:QVAE} we introduce QVAEs and present our results.  We conclude in Sec.~\ref{sec:conc} and give further technical and methodological details in the Appendices.

\section{Generative Models with latent Variables}
\label{sec:GEN}

Let $X = \{  \x^{d}\}_{d=1}^N$ represents a training set of $N$ independent and identically distributed samples coming from an unknown data distribution, $p_\text{data}(X)$  (for instance, the distribution of the pixels of a set of images). Generative models are probabilistic models that minimize the ``distance" between the model distribution, $p_\btheta(X)$, and the data distribution, $p_\text{data}(X)$, where $\btheta$ denotes the parameters of the model. Generative models can be categorized in several ways, but for the purpose of this paper we focus on the distinction between models with latent (unobserved) variables and fully-visible models with no latent variables. Examples of the former include generative adversarial networks (GAN)~\cite{goodfellow_generative_2014}, VAEs~\cite{kingma2013auto}, and RBMs, whereas some important examples of the latter include NADE~\cite{uria2016neural}, MADE~\cite{germain2015made}, pixelRNNs, and pixelCNNs~\cite{oord2016pixel}.

\begin{figure*}[t]
\begin{center}
\subfigure[\, Undirected generative model with latent variables]{\includegraphics[, width=.55\columnwidth]{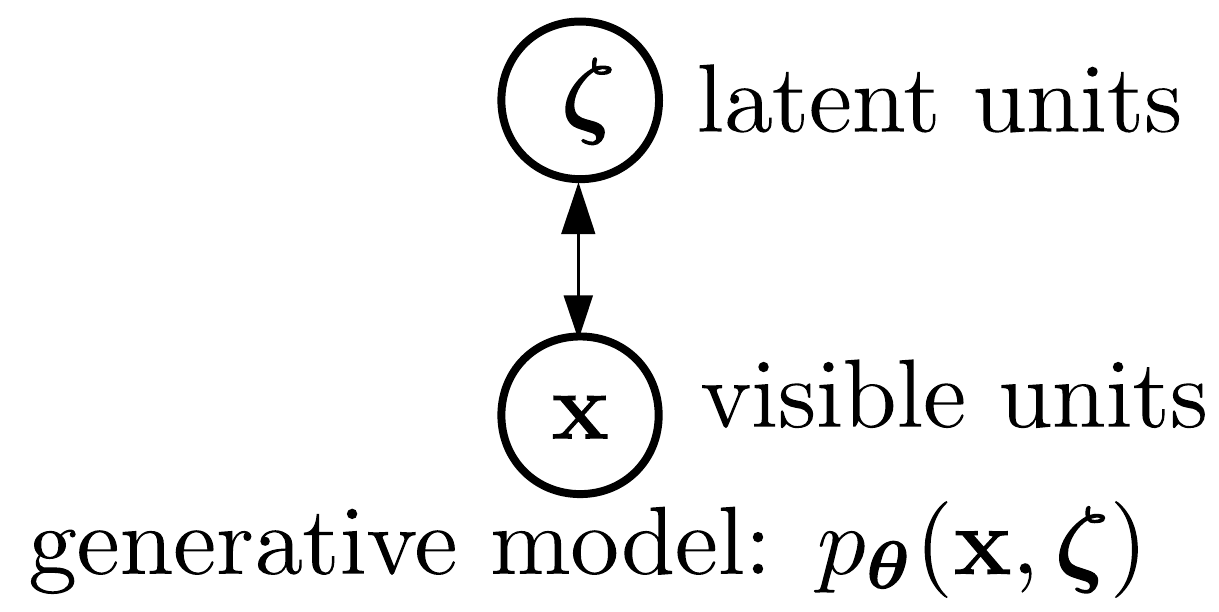}\label{fig:schema}}
\subfigure[\, Directed generative model (VAE)]{\includegraphics[, width=1\columnwidth]{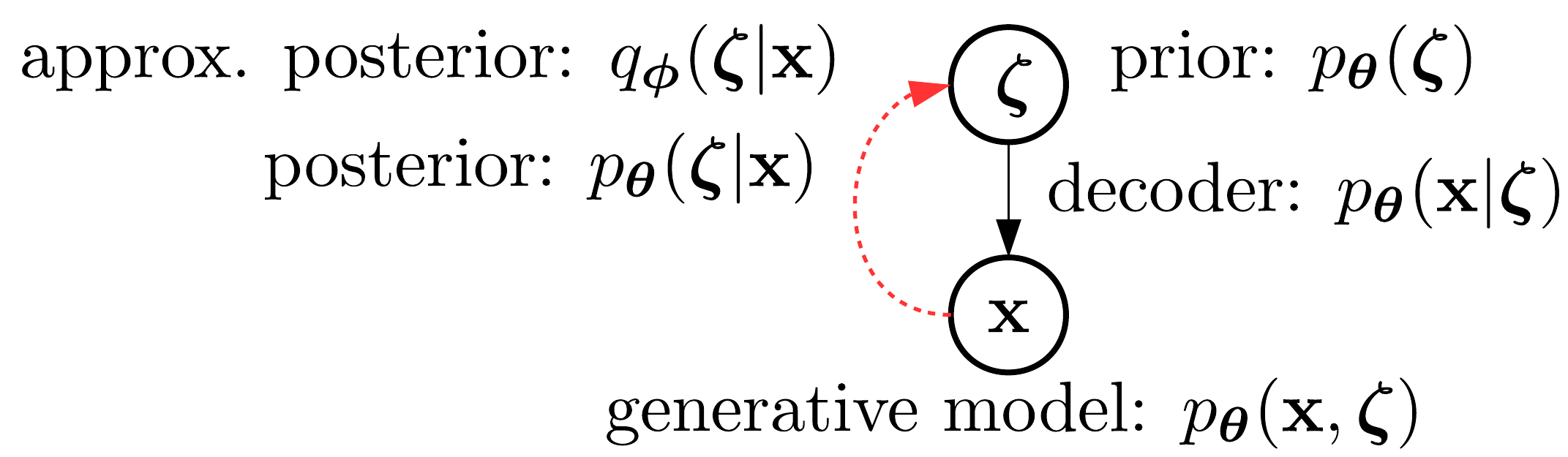}\label{fig:schemb}}
\subfigure[\, DVAE or QVAE]{\includegraphics[, width=.5\columnwidth]{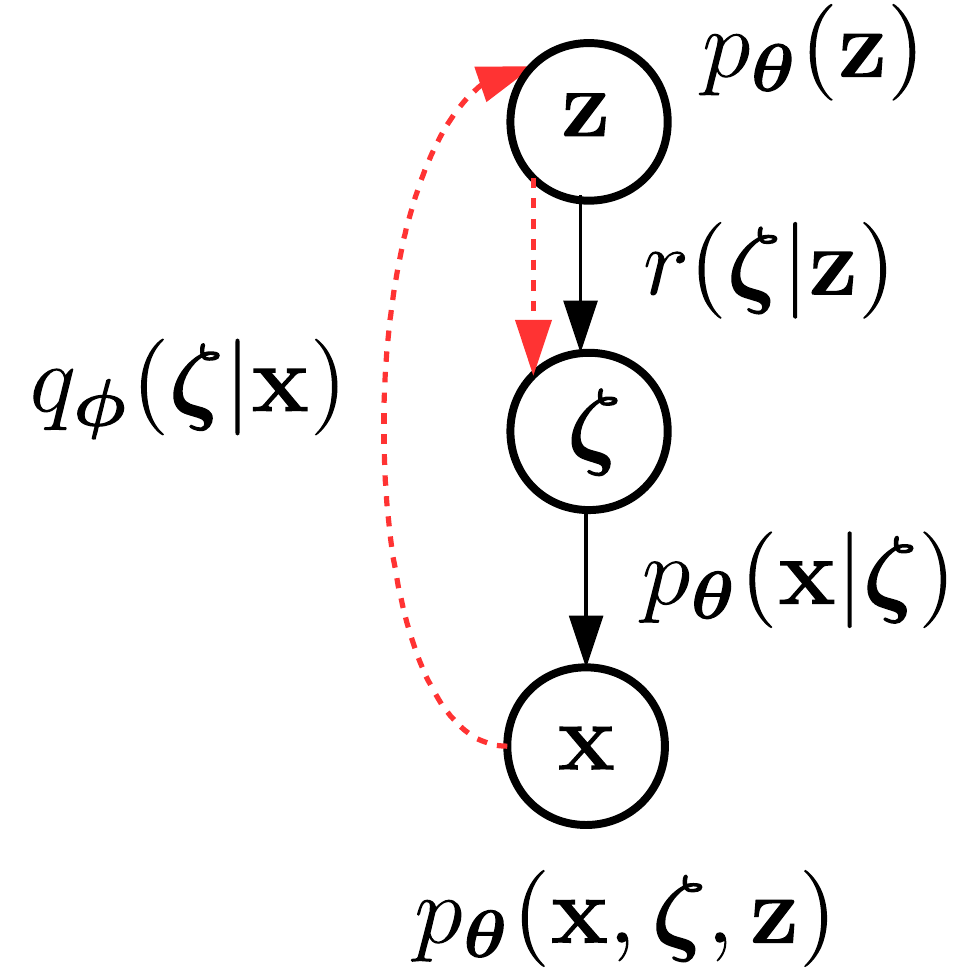}\label{fig:schemc}}
\caption{Generative models with latent variables can be represented as graphical models that describe conditional relationships within variables. a) Undirected generative models are defined in terms of the a joint probability distribution, $p_{\btheta}(\x, \bzeta)$. Boltzmann machines belong to this group of generative models. b) In a directed generative model, the joint probability distribution  $p_{\btheta}(\x, \bzeta)$, is decomposed as $p_{\btheta}(\x, \bzeta) = p_{\btheta}(\x|\bzeta) p_{\btheta}(\bzeta)$: a prior distribution over the latent variables  $p_{\btheta} (\bzeta )$ and a decoder distribution $p_{\btheta} (\x|\bzeta )$. The prior and the decoder are ``hard-coded" or explicitly determined by the model; however, the posterior,  $p_{\btheta}(\bzeta| \x)$ (dotted red arrow) is intractable. In VAEs, an approximating posterior, $q_{\bphi}(\bzeta|\x)$, is proposed to replace the intractable true posterior. c) Structure of the generative and the inference (red dotted arrows) models of a DVAE and a QVAE. Here, $p_{\btheta} (\z)$ represents the prior over discrete variables, $\z$, and is characterized by an RBM or a QBM in DVAEs or QVAEs, respectively. The continuous variables $\bzeta$ are introduced to allow for a smooth propagation of the gradients.
}
\label{fig:schem}
\end{center}
\end{figure*}

The conditional relationships among the visible units, $\x$, and latent units, $\bzeta$, determine the joint probability distribution, $p_{\btheta}(\x, \bzeta)$,  of a generative model and can be represented in terms of either \emph{undirected}  (Fig.~\ref{fig:schema}) or \emph{directed} (Fig.~\ref{fig:schemb}) graphs.
Unlike fully visible models, generative models with latent variables can potentially learn and encode in the latent space useful representations of the data. This is an appealing property that can be exploited to improve other tasks such as supervised and semi-supervised  learning (\emph{i.e.}, when only a fraction of the input data is labeled)~\cite{kingma2014semi} with substantial practicality in image search~\cite{fergus2009semi}, speech analysis~\cite{liu2013graph}, genomics~\cite{shi2011semi}, drug design~\cite{chen2013semi}, and so on.

Training a generative model is commonly done via maximum likelihood (ML) approach, in which optimal model parameters $\btheta^*$ are obtained by maximizing the likelihood of the dataset:
\beq
\sum_{\bf x} p_\text{data}(\x) \log p_{\btheta} (\x) = \Ex{[\log p_{\btheta}(\x)]}\,,
\eeq
where $ p_{\btheta} (\x) = \sum_{\bzeta}  p_{\btheta} (\x,\bzeta)$ is the marginal probability distribution of the visible units and $\Ex[\dots]$ means the expectation value over $\x$ sampled from $p_\text{data}(\x)$.

To better understand the behavior of generative models with latent variables, we now write $\Ex[{\log p_{\btheta}(\x)}]$ in a more insightful form. First, note that $\log p_{\btheta}(\x)= \mathbb{E}_{\bzeta \sim p_{\btheta} (\bzeta | \x)} [\log p_{\btheta}(\x)]$, since $p_{\btheta}(\x)$ is independent of $\bzeta$. The quantity $p_{\btheta} (\bzeta | \x)$ is called the \emph{posterior} distribution, since it represents the probability of the latent variables after an observation $\x$ has been made (see Fig.~\ref{fig:schemb}). Also, since we have $p_{\btheta}(\x) = {p_{\btheta}(\x, \bzeta)}/{p_{\btheta} (\bzeta | \x)} $, we can write:
\bea
\Ex{[\log p_{\btheta}(\x)]} = \ph\ph\ph\ph\nn
  =  \Ex\left[\mathbb{E}_{\bzeta \sim p_{\btheta} (\bzeta | \x)}\left[{\log \frac{p_{\btheta}(\x, \bzeta)}{p_{\btheta} (\bzeta | \x)}}\right]\right]\,.
\label{eq: logp_gen}
\eea
By noticing that $p_{\btheta} (\bzeta,\x) = p_{\btheta} (\bzeta) p_{\btheta} (\x| \bzeta)$ and rearranging Eq.~\ref{eq: logp_gen}, we have:
\bea
\Ex{[\log p_{\btheta}(\x)]} =\Ex \Bigg[\mathbb{E}_{\bzeta \sim p_{\btheta} (\bzeta | \x)}
[{\log p_{\btheta} (\x| \bzeta)}] \nn
 - \underbrace{ 
\mathbb{E}_{\bzeta \sim p_{\btheta} (\bzeta | \x)} 
\Bigg[\log \frac{p_{\btheta} (\bzeta | \x)}{p_{\btheta} (\bzeta )}
}_{D_{KL}(p_{\btheta} (\bzeta | \x)||p_{\btheta} (\bzeta ))}\Bigg]\Bigg],\ph
\label{eq: logp_elbo_like}
\eea
where $ p_{\btheta} (\bzeta) = \sum_{\x}  p_{\btheta} (\x,\bzeta)$ is the \emph{prior} distribution. The term $D_{KL}(p||q) \equiv{\mathbb{E}_{p} \log [{p}/{q} ]}$ represents the Kullback-Leibler (KL) divergence, which is a measure of ``distance" between the two distributions $p$ and $q$~\cite{bishop_pattern_2011}. 

Maximizing the first term maximizes the probability of $p_{\btheta} (\x | \bzeta)$ when $\bzeta$ is sampled from $p_{\btheta} (\bzeta | \x)$ for a given input from the dataset. This is called reconstruction, because it implies that samples from  $p_{\btheta} (\x | \bzeta)$ have maximum similarity to the input $\x$.
This is an ``autoencoding" process and hence the first term in Eq.~(\ref{eq: logp_elbo_like}) is the autoencoding term.
Conversely, maximizing the second term, corresponds to minimizing the expected KL divergence under $p_\text{data}$. For a given input $\x$, this amounts to minimizing the distance between the posterior $p_{\btheta} (\bzeta | \x)$ and the prior $p_{\btheta} (\bzeta )$. In the limiting case, this leads to $p_{\btheta} (\bzeta | \x)=p_{\btheta} (\bzeta )$, which is only possible if $p_{\btheta} (\x,\bzeta)=p_{\btheta} (\bzeta )p_{\btheta} (\x )$. Therefore, the \emph{mutual information}~\cite{cover_elements_1991} between $\x$ and $\bzeta$ is zero. In other words, while the autoencoding term strives to maximize the mutual information, the KL term tries to minimize it.


Eventually, the amount of information condensed in the latent space depends on the intricate balance between the two terms in Eq.~\ref{eq: logp_elbo_like}, which in turn depends on the type of generative model chosen and also on the training method used. For example, directed models such as those depicted in Fig.~\ref{fig:schemb} are characterized in terms of explicitly defining the prior $p_{\btheta} (\bzeta)$ and the decoder $p_{\btheta} (\x |\bzeta)$ distributions. If the decoder distribution has high representation power, it can easily decouple $\x$ and $\bzeta$ to avoid paying the KL penalty. This leads to poor reconstruction quality. On the other hand, if the decoder is less expressive (as is a neural net yielding the parameters of a factorial Bernoulli distribution) a high amount of information is stored in the latent space and the model autoencodes to a good degree.

\section{Variational Autoencoders}
\label{sec:VAE}

A common problem of generative models with latent variables is the intractability of \emph{inference}, \emph{i.e.}, calculating the posterior distribution $p_{\btheta}(\bzeta|\x) = p_{\btheta}(\x |\bzeta) p_{\btheta}(\bzeta)/{p_{\btheta} (\x)}$. This involves the evaluation of
\beq
p_{\btheta}(\x) = \int p_{\btheta}(\x|\bzeta)p_{\btheta}(\bzeta) d\bzeta\,.
\label{eq:px}
\eeq
 The first crucial element of the VAE setup is variational inference; \emph{i.e.}, introducing a tractable variational approximation $q_{\bphi}(\bzeta|\x)$ (Fig.~\ref{fig:schemb}) to the true posterior $p_{\btheta}(\bzeta|\x)$~\cite{hoffman2013stochastic}, with variational parameters $\bphi$. 
 Both decoder $p_{\btheta}(\x|\bzeta)$ and encoder $q_{\bphi}(\bzeta|\x)$ are commonly implemented by neural networks, known as generative and recognition (inference) networks, respectively.
 
 To define an objective function for optimizing parameters $\btheta$ and $\bphi$, we can replace  $p_{\btheta}(\bzeta|\x)$ with $q_{\bphi}(\bzeta|\x)$ in Eq.~\ref{eq: logp_elbo_like}:
\bea
\mathcal L(\btheta,\bphi) & \equiv &  \Ex[ \mathcal L(\btheta,\bphi, \x)] \equiv \nn
 & = &  \Ex[  \E_{\bzeta\sim q_{\bphi}(\bzeta|\x)}[\log p_{\btheta}(\x|\bzeta)]  + \nn
&- & D_{KL}( q_{\bphi}(\bzeta|\x) ||  p_{\btheta}(\bzeta))]\,.
\label{eq: ELBO}
\eea 
Although $\mathcal L(\btheta,\bphi)$ is not equal to the log-likelihood, it provides a lower bound: 
\bea 
\mathcal L(\btheta,\bphi) \le  \Ex[\log p_{\btheta}(\x)]\,,
\label{eq: elbo_ge}
\eea
as we show below. 
Because of this important property, $\mathcal L(\btheta,\bphi)$ is called the evidence (variational) lower bound (ELBO).  To prove Eq.~\ref{eq: elbo_ge}, we note from Eq.~\ref{eq: ELBO} 
that 
\bea
\mathcal L(\btheta,\bphi, \x) &=& \E_{\bzeta\sim q_{\bphi}(\bzeta|\x)}
\left[\log p_{\btheta}(\x|\bzeta)- \log \frac{q_{\bphi}(\bzeta|\x)}{p_{\btheta}(\bzeta)}\right]= \nn
&=& \E_{\bzeta\sim q_{\bphi}(\bzeta|\x)}\left[ \log \frac{p_{\btheta}(\x, \bzeta)}{q_{\bphi}(\bzeta|\x)}\right] \label{eq:elbo_undivided}
\eea
where we have used $p_{\btheta} (\x,\bzeta) = p_{\btheta} (\bzeta) p_{\btheta} (\x| \bzeta)$. Eq.~\ref{eq:elbo_undivided} is a compact way of expressing the ELBO, which will be used later. One may further use $p_{\btheta} (\x,\bzeta) = p_{\btheta} (\x) p_{\btheta} (\bzeta|\x)$ to obtain yet another way of writing the ELBO:
\bea
\mathcal L(\btheta,\bphi, \x) 
&=& \log p_{\btheta}(\x) - \E_{\bzeta\sim q_{\bphi}(\bzeta|\x)}\left[ \log \frac{q_{\bphi}(\bzeta|\x)}{p_{\btheta}(\bzeta|\x)}\right]\nn 
&=& \log p_{\btheta}(\x) - D_{KL}( q_{\bphi}(\bzeta|\x) ||  p_{\btheta}(\bzeta|\x))]. \label{ELBOKL}
\eea
Since KL divergence is always non-negative, we obtain
\bea
\mathcal L(\btheta,\bphi, \x) \le \log p_{\btheta}(\x),
\eea
which immediately gives Eq.~\ref{eq: elbo_ge}. 

It is evident from Eq.~\ref{ELBOKL} that the difference between the ELBO and the true log-likelihood, \emph{i.e.}, the tightness of the bound, depends on the distance between the approximate and true posteriors. Maximizing the ELBO, therefore, increases the log-likelihood and decreases the distance between the two posterior distributions at the same time. Success in minimizing the bound between the log-likelihood and ELBO depends on the flexibility and representational power of $q_{\bphi}(\bzeta|\x)$. However, increasing the representational power of $q_{\bphi}(\bzeta|\x)$ does not guarantee success in encoding the information in the latent space. In other words, the widespread problem~\cite{chen2016variational, zhao2017infovae, yeung2017tackling, tomczak2017vae, zhao2017towards} of ``ignoring the latent code" in VAEs is not \emph{completely} an artifact of choosing a family of approximating posterior distributions with limited representational power. As we discussed before, it is rather an intrinsic feature of generative models with latent variables due to the clash of the two terms in the objective function defined in Eq.~\ref{eq: logp_elbo_like}. 


\subsection{The reparameterization trick}

The objective function in Eq.~\ref{eq:elbo_undivided} contains expectation values of functions of the latent variables $\bzeta$ under the posterior distribution $q_{\bphi}(\bzeta|\x)$. To train the model, we need to calculate the derivatives of these terms with respect to $\btheta$ and $\bphi$. However, evaluating the derivatives  with respect to $\bphi$ is problematic  because the expectations of Eq.~\ref{eq:elbo_undivided} are estimated using samples that are generated  according to a probability distribution that depends on  $\bphi$. A naive solution to the problem of calculating $\pd_{\bphi}$ of the expected value of an arbitrary function $ \E_{\bzeta \sim q_\phi}[f(\bzeta)]$, is to use the identity $\pd_{\bphi} q_\phi = q_\phi \pd_{\bphi} \log q_\phi $, to write
\bea
\pd_{\bphi} \E_{\bzeta \sim q_\phi}\left[f(\bzeta)\right] = \E_{\bzeta \sim q_\phi}\left[f(\bzeta) \pd_{\bphi} \log q_\phi \right]\,.
\label{eq:REINFORCE}
\eea
Here, for simplicity we assumed that $f$ does not depend on $\bphi$. This approach is known as the REINFORCE. However, the expectation of Eq.~\ref{eq:REINFORCE} has high variance and requires intricate variance-reduction mechanisms to be of practical use~\cite{mnih2014neural}.

A better approach is to write the random variable $\bzeta$ as a deterministic function of the distribution parameters $\bphi$ and of an additional auxiliary random variable $\brho$. The latter is given by a probability distribution $p(\brho)$ 
that does not depend on $\bphi$. This reparameterization, $\bzeta(\bphi , \brho)$, can be used to write $\E_{\bzeta \sim q_\phi}[f(\bzeta)] = \E_{\brho \sim p(\brho)}[f( \bzeta(\bphi, \brho))]$. Therefore, we can move the derivative inside the expectation with no difficulty:
\bea
\pd_{\bphi} \E_{\bzeta \sim q_\phi}\left[f(\bzeta)\right] = \E_{\brho \sim p(\brho)}\left[ \pd_{\bphi} f( \bzeta(\bphi, \brho)) \right]. \label{eq:reparam_preamble}
\eea
This is called the \emph{reparameterization  trick}~\cite{kingma2013auto} and is 
mostly responsible for the recent success and proliferation of VAEs. When applied to Eq.~\ref{eq:elbo_undivided}, we have:
\bea
\mathcal L(\btheta,\bphi, \x) &=& \E_{\bzeta\sim q_{\bphi}(\bzeta|\x)}\left[ \log \frac{p_{\btheta}(\x, \bzeta)}{q_{\bphi}(\bzeta|\x)}\right] \nn
&=&
\E_{\brho \sim p(\brho)}\left[ \log \frac{p_{\btheta}(\x, \bzeta(\bphi, \brho))}{q_{\bphi}(\bzeta(\bphi, \brho)|\x)}\right],
\label{eq:elbo_reparam_applied}
\eea
where we have suppressed the inclusion of $\x$ in the arguments of the reparameterized $\bzeta$ to keep the notation uncluttered.

It is now important to find a function $\bzeta(\bphi,\brho)$ such that $\brho$ becomes $\bphi$-independent. Let us define a function $\mathbf F$
\beq
\brho \equiv \mathbf F_\bphi(\bzeta).
\eeq 
The probability distributions $p(\brho)$ and $q_\bphi(\bzeta|\x)$ should satisfy $p(\brho)d\brho = q_\bphi(\bzeta|\x) d\bzeta$, therefore
\beq
p(\brho) = {q_\bphi(\bzeta|\x) \over d\brho/d\bzeta} = {q_\bphi(\bzeta|\x) \over d \mathbf F_\bphi(\bzeta)/ d\bzeta}. \label{Eq:CDFproof}
\eeq
To have $p(\brho)$ independent of $\bphi$ we need 
\beq
\mathbf F_\bphi(\bzeta) = \int_0^{\bzeta} q_\bphi(\bzeta'|\x) d\bzeta'.
\eeq
Now, by choosing $\mathbf F_\bphi$ to be the cumulative distribution function (CDF) of $q_\bphi(\bzeta|\x)$, $p(\brho)$ becomes a uniform distribution $\mathcal{U}(0,1)$ for $\brho \in [0,1]$. We can thus write
\beq
\bzeta(\bphi,\brho) = F^{-1}_\bphi(\brho)\,.
\label{eq:inverse_CDF}
\eeq
To derive Eq.~\ref{eq:inverse_CDF}, we have implicitly assumed that the latent variables are continuous and that the posterior factorizes:  $q_\bphi(\bzeta|\x) = \prod_l q_\bphi(\zeta_l|\x)$. It is possible to extend the reparameterization trick to include discrete latent variables (see  next section) and more complicated approximate posteriors (see Appendix~\ref{sec:HIER}).

\section{VAE with discrete latent space}
\label{sec:DVAE}
Most of the VAEs studied so far have continuous latent spaces due to the difficulty of propagating derivatives through discrete variables. Nonetheless, discrete stochastic units are indispensable to representing distributions in supervised and unsupervised learning, attention models, language modeling and reinforcement learning~\cite{jang2016categorical}. Some noteworthy examples include application of discrete units in learning distinct semantic classes~\cite{kingma2014semi} and in semisupervised generation~\cite{maaloe2017semi} to learn more meaningful hierarchical VAEs. In Ref.~\cite{makhzani2017pixelgan}, when the latent space is composed of discrete variables, the representations learn to disentangle content and style information of images in an unsupervised fashion.  

Due to the non-differentiability of discrete stochastic units, several methods that involve variational inference  use the REINFORCE method, Eq.~\ref{eq:REINFORCE}, from the reinforcement learning literature~\cite{paisley2012variational, mnih2014neural, gu2015muprop}. However, these methods yield noisy estimates of gradients that need to be mitigated using several variance reduction techniques such as finding appropriate control variates. Another approach involves using biased derivatives for the Bernoulli variables~\cite{bengio2013estimating}.  There are also two approaches that extend the reparameterization trick to discrete variables. Refs.~\cite{jang2016categorical, maddison2016concrete} concurrently came up with a relaxation of categorical discrete units into continuous variables by adding Gumbel noise to the logits inside a softmax function, with a temperature hyper-parameter. The softmax function transforms into a non-differentiable argmax function obtaining unbiased samples in the limit of zero temperature. However, in this limit the training stops since variables become truly discrete. Therefore, an annealing schedule is used for the temperature throughout the training to obtain less noisy, yet biased, estimates of gradients~\cite{jang2016categorical}. 

Here we follow the approach proposed in~\cite{rolfe2016discrete}, which yields reparameterizable and unbiased estimates of gradients. As discussed in the previous section, the generative process in a VAE involves sampling a set of continuous variables $\bzeta \sim p_{\btheta}(\bzeta)$. To implement a DVAE, we assume the prior distribution is now defined on a set of discrete variables  $\z \sim p_{\btheta}(\z)$, with $\z \in \{0, 1 \}^L$. Once again we use $\btheta$ to denote collective parameters of the generative side of the model. To propagate the gradients through the discrete variables, we keep the variables $\bzeta$ as an auxiliary set of continuous variables~\cite{rolfe2016discrete}. The full prior is chosen as follows (Fig.~\ref{fig:schemc}) :
\beq
p_\btheta(\bzeta,\z) \equiv r(\bzeta|\z)p_\btheta(\z) \equiv  \left( \prod_{l=1}^L r(\zeta_l|z_l) \right) p_\btheta(\z)\,.
\eeq
The newly introduced term $r(\bzeta|\z)$ acts as a smoothing probability distribution that enables the implementation of the reparameterization trick. The structure of the DVAE is completed by considering a particular form for the approximating posterior and marginal distributions (Fig.~\ref{fig:schemc}): 
\bea
q_\bphi(\bzeta, \z|\x) & \equiv&  r(\bzeta|\z)q_\bphi(\z|\x) \nn
 p_\btheta(\x|\bzeta,\z) & \equiv & p_\btheta(\x|\bzeta)\,,
\eea
where for now we assume $q_\bphi(\z|\x)=  \prod_l q_\bphi(z_l|\x)$ is a product of Bernoulli probabilities  for the discrete variable $z_l$ (see again Appendix~\ref{sec:HIER} for the case where hierarchies are present in the posterior). With the above choice, the ELBO bound can be written as
\bea
\mathcal L(\btheta,\bphi, \x) & = &    \E_{q_{\bphi}(\bzeta|\x)}[\log p_{\btheta}(\x|\bzeta)] + \nn
& - & D_{KL}( q_{\bphi}(\z|\x) ||  p_{\btheta}(\z))\,,
\label{eq:D-ELBO}
\eea 
where $q_\bphi(\bzeta|\x)$ is the approximate posterior marginalized over the discrete variables. In the equation above we have used the fact that the KL term does not explicitly depend on $\bzeta$ while the autoencoding term does not explicitly depend on $\z$. 

\subsection{The reparameterization trick for DVAE}
\label{sec:dvae_repar}

We can  apply the inverse CDF reparameterization trick, Eq.~\ref{eq:inverse_CDF}, to the autoencoding term in Eq.~\ref{eq:D-ELBO} if we choose the function $ r(\bzeta|\z)$ such that the CDF of the approximating posterior marginalized over the discrete variables
\beq
 {\rm \mathbf F}(\bzeta) \equiv \int_{0}^{\bzeta} q_{\bphi}(\bzeta' |\x) d \bzeta'  
\eeq
 can be inverted:
\bea
 \E_{q_{\bphi}(\bzeta|\x)}[\log p_{\btheta}(\x|\bzeta)]  =  \E_{\brho \sim \mathcal{U}}[\log p_{\btheta}(\x|{\rm \mathbf F}^{-1}(\brho)]\,. 
 \label{eq:trick}
\eea

An appropriate choice for $r(\zeta_l|z_l)$ is, for example, the spike-and-exponential transformation:
\bea
r(\zeta_l|z_l=0) & = & \delta(\zeta_l) \nonumber \\
r(\zeta_l|z_l= 1)  &=& \left\{ 
\begin{array}{ll}
\beta \frac{ e^{\beta \zeta_l}}{e^\beta -1}, & {\rm if } \quad 0 < \zeta_l \le 1   \\
   0, & {\rm otherwise}\,. 
\end{array}
\right. \label{Spike}
\eea
For this distribution we can write:
\beq
{\rm F}_l(\zeta_l) = \int_0^{\zeta_l} q_\bphi(\zeta'_l|\x) d\zeta'_l = \int_0^{\zeta_l} \sum_{z_l=0,1}q_\bphi(z_l|\x)r(\zeta'_l|z_l)d\zeta'_l\,.
\eeq
Using Eq.~\ref{Spike} with Bernoulli distribution $q_\bphi(z_l{=}1|\x)=q_l$ and $q_\bphi(z_l{=}0|\x)=1{-}q_l$, we find
\beq
\rho_l = q_l \frac{e^{\beta \zeta_l}-1}{e^\beta-1}  + (1-q_l)\,,
\eeq
which  can be easily inverted to obtain $\zeta_l$
\bea
\zeta_l (\rho_l,q_l)  &=&  \frac1\beta \log\left[ \left(\frac{{ \rm max}(\rho_l{+}q_l{-}1, 0)}{q_l} \right)(e^\beta{-}1) + 1  \right]\,. \nonumber \\
\label{eq:F}
\eea
The virtue of the spike-and-exponential smoothing distribution is that $z_l$ can be deterministically obtained from $\zeta_l$ and thus $\rho_l$:
\beq
z_l(\rho_l,q_l) = {\rm sign}(\zeta_l(\rho_l,q_l)) = \Theta(\rho_l{+}q_l{-}1)\,,
\label{eq:z_given_zeta}
\eeq
which follows from Eqs~\ref{Spike} and~\ref{eq:F}. This property is crucial to apply the reparameterization trick to the KL term, as shown below, and evaluating its derivatives as shown in Appendix~\ref{sec:dervs}.

For later convenience, we note that the KL term can be written as the difference between an \emph{entropy} term, $ H(q_{\bphi}(\z|\x))$, and a \emph{cross-entropy} term, $H(q_{\bphi}(\z|\x), p_{\btheta}(\z))$:
\bea
D_{KL}(q_{\bphi}(\z|\x)||p_{\btheta}(\z)) =
\underbrace{\mathbb{E}_{q_{\bphi}}[\log q_{\bphi}]}_{- H(q_{\bphi})} - \underbrace{\mathbb{E}_{q_\bphi}[\log p_{\btheta}]}_{-H(q_{\bphi}, p_{\btheta} ) }\,. \nonumber
\eea
Herein, for simplicity, we use $q_{\phi}$ and $p_{\theta}$ in place of $q_{\bphi}(\z|\x)$ and $p_{\btheta}(\z)$, respectively, in unambiguous cases. Using Eq.~\ref{eq:z_given_zeta}, the reparameterization trick can be applied to the entropy term:
\bea
H(q_\bphi) \equiv  -  \E_{\z \sim q_\bphi}[\log q_\bphi] =
 -  \E_{\brho \sim \mathcal{U}}[\log q_\bphi (\z(\brho, \bphi)|\x)],\nn
 \eea
where we have explicitly shown the dependence of $\z$ on $\brho$ and $\bphi$.
Note that in the simple case of a factorial Bernoulli distribution, we do not need to use the reparameterization trick and can use the analytic form of the entropy; \emph{i.e.}, $H(q_\bphi) = - \sum_{l=1}^L\left(  q_l \log q_l + (1-q_l) \log (1-q_l)  \right)$ (see Appendix~\ref{sec:BERN} for more details). Similarly, applying the reparameterization trick to the cross-entropy leads to:
\beq
-H(q_\bphi, p_\btheta ) \equiv   \E_{\z \sim q_\bphi}[\log p_\btheta] = \E_{\brho \sim \mathcal{U}}[\log p_\btheta(\z(\brho, \bphi))]\,.
\eeq

It is a common practice to use hierarchical distributions to achieve more powerful approximating posteriors. Briefly, the latent variables are compartmentalized into several groups, and the probability density function of each group depends on the values of the latent variables in the preceding groups; \emph{i.e.}, $q_{\bphi}(z_l|\zeta_{m<l}, \x)$. This creates a more powerful approximating posterior able to represent more complex correlations between latent variables, as compared to a simple factorial distribution. See Appendix~\ref{sec:HIER} for more details.

\subsection{DVAE with Boltzmann machines}

Boltzmann machines are probabilistic models able to represent complex multi-modal probability distributions~\cite{ackley1985learning}, and are thus attractive candidates for the latent space of a VAE. 
This approach is also appealing with regards to the machine-learning application of quantum computers. The probability distribution realized by an RBM is
\bea
p_\btheta(\z) & \equiv &    {e^{-E_\btheta(\z)}}/Z_\btheta\,, \quad Z_\btheta \equiv {\sum_\z e^{-E_\btheta(\z)}}   \,,   \nn
 E_\btheta(\z)& =&  \sum_l z_l h_l + \sum_{l< m}  W_{lm} z_l z_m, \quad {\bf h}, {\bf W} \in \{\btheta\}.\quad
\eea
The negative cross entropy term $-H(q_\bphi, p_\btheta )=\E_{\z \sim q_\bphi}[\log p_\btheta]$ is the log-likelihood of $\z$ sampled from the approximating posterior  $\z \sim q_\bphi(\z|\x)$ under the model $p_\btheta$.
After reparameterization, we have
\bea
H(q_\bphi, p_\btheta ) &=&  - \E_{\brho \sim \mathcal{U}}[\log p_\btheta(\z(\brho, \bphi))] \nn
&=&    \E_{\brho \sim \mathcal{U}}[ E_\btheta(\z(\brho, \bphi))] + \log Z_\btheta.
\eea
Gradients can thus be computed as usual as the difference between a positive and negative phase, in which the latter is computed via Boltzmann sampling from the BM:
\beq
\pd H(q_\bphi, p_\btheta ) = \E_{\brho \sim \mathcal{U}}[\pd E_\btheta(\z(\brho,\bphi))] - \E_{\z \sim p_{\btheta}}[{\pd E_\btheta(\z)}]\,.
\label{eq:posneg}
\eeq

Notice that the positive phase (the first term above) involves the expectation over the approximating posterior, but it is explicitly written in terms of the discrete variables $\z(\brho,\bphi)$. We thus need to calculate the derivatives through these variables. We discuss the computation of the positive phase in the most general case in Appendix~\ref{sec:dervs}.

\subsection{Experimental results with DVAE}

In this section, we show that the DVAE model introduced in Sec.~\ref{sec:DVAE} achieves state-of-the-art performance, for variational inference models with only latent variables, on the MNIST dataset~\cite{lecun1998mnist}.  We perform experiments with restricted Boltzmann machines, in which the hidden and visible units are placed at the two sides of a bipartite graph. Notice that in a DVAE setup, all the units of the (classical) RBM are latent variables (there is technically no distinction between visible and hidden units as for standalone RBMs). We still use an RBM to exploit its bipartite structure enabling efficient Gibbs block-sampling.  This allows us to train DVAEs with RBMs with up to $256$ units per layer. 

\begin{figure}[t]
\begin{center}
\subfigure[\, Generated MNIST: DVAE, RBM$_{32\times 32}$]{\includegraphics[width=0.49\columnwidth]{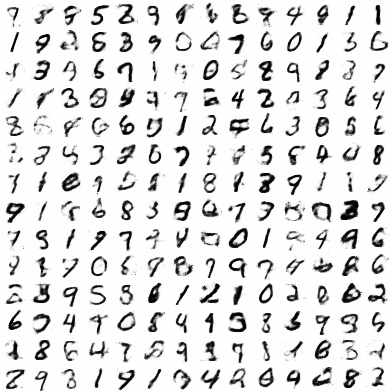}\label{fig:1a}}
\subfigure[\, Reconstructed MNIST: DVAE, RBM$_{32\times 32}$]{\includegraphics[width=0.49\columnwidth]{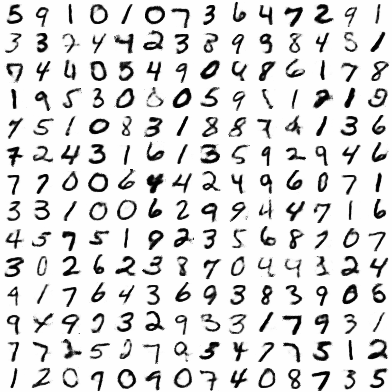}\label{fig:1b}}\\
\subfigure[\, Generated MNIST: DVAE, RBM$_{256\times 256}$]{\includegraphics[width=0.49\columnwidth]{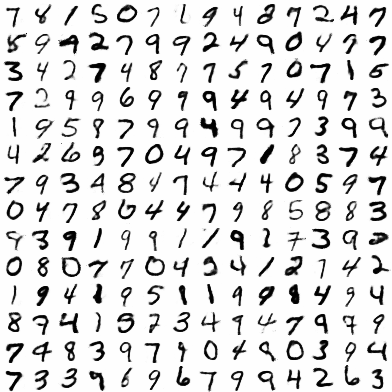}\label{fig:1c}}
\subfigure[\, Reconstructed MNIST: DVAE, RBM$_{256\times 256}$]{\includegraphics[width=0.49\columnwidth]{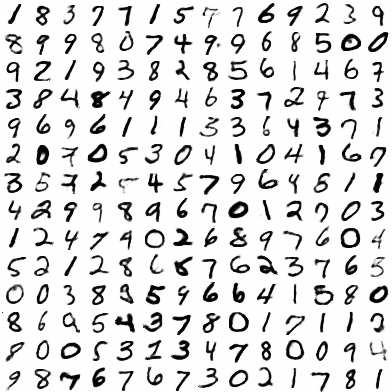}\label{fig:1d}}
\caption{Generated and reconstructed MNIST with DVAE.}
\label{fig:1}
\end{center}
\end{figure}

Figure~\ref{fig:1} shows generated and reconstructed MNIST digits for a DVAE with RBMs with 32 and 256 units per layer.  In Table~\ref{tab:1}, we report the best results for the ELBO and log-likelihood (LL) we obtained with RBMs  of $32, 64, 128$, and $256$  units per layer.  For $256$ units, we obtained an LL of $-83.5 \pm 0.2$, with the reported error being a conservative estimate of our statistical uncertainty. In all cases, the negative phase of the RBMs was estimated using  persistent contrastive divergence (PCD), with $1000$ chains and $200$ block-Gibbs updates per gradient evaluation. We have chosen an approximating posterior with 8 levels of hierarchies (the number of units that each level of hierarchy represents is the total number of latent units divided by 8);  each Bernoulli probability $q_{\bphi}(z_l|\zeta_{m<l}, \x)$ is a sigmoidal output of a feed-forward neural network with two hidden rectified linear unit (ReLU) layers containing $2000$ deterministic units.  

The model is prone to overfitting when representing the decoder distribution $p_\btheta(\x|\bzeta)$ with deep networks. We considered $p_\btheta(\x|\bzeta)$ to be sigmoidal outputs of a ReLu network with one layer and the number of deterministic units that vary between $250$ and $2000$. Typically, a larger RBM required a smaller number of hidden units in the decoder network to avoid overfitting. Our implementation included annealing schedules for both the learning rate (exponential decay) and the $\beta$ parameter (linear increase) in Eq.~\ref{Spike}. Batch normalization~\cite{ioffe2015batch} was used to expedite the training process. The value of $\beta$ was annealed throughout the training from 1.0 to 10 during 2000 epochs with a batch size of 200. We used the ADAM stochastic optimization method with a learning rate of $10^{-3}$ and the default extra parameters~\cite{kingma2014adam}.  To calculate the LL in Table~\ref{tab:1}, we used importance weighting to compute a  {multi-sample ELBO,  as delineated in~\cite{burda2015importance}, with $30000$ samples in the latent space for each input image in the test set. It can be shown that the value of the multi-sample ELBO asymptotically reaches the true LL when the number of samples approaches infinity~\cite{burda2015importance}. The $\log Z$ was computed using population annealing~\cite{hukushima2003population, machta2010population} (see also Appendix~\ref{sec:ctqmc} for the quantum partition function). In all our experiments we have verified that the statistical error on the evaluation of $\log Z$ is negligible.

In Table~\ref{tab:1}, we also report the results of some other algorithms that use discrete variables in variational inference. NVIL~\cite{mnih2014neural} and its importance weighted analog, VIMCO~\cite{mnih2016variational}, use the REINFORCE trick, Eq.~\ref{eq:REINFORCE} along with carefully designed control variates to reduce the variance of the estimation. CONCRETE~\cite{maddison2016concrete} and Gumbel-Softmax ~\cite{jang2016categorical} are two concurrently developed methods that are based on applying the reparameterization trick to discrete latent variables. RWS~\cite{bornschein2014reweighted} is a multi-sampled and improved version of the wake-sleep algorithm~\cite{hinton1995wake}, which can be considered as a variational approximation (since an encoder or ``inference network" is present) with different loss functions in the wake and sleep phases of training. REBAR~\cite{tucker2017rebar} is the application of the CONCRETE method to create control variates for the REINFORCE approach. All algorithms reported in  Table~\ref{tab:1}, excluding DVAE, implement a latent space with independent discrete units distributed according to a set of independent Bernoulli distributions. The result reported for CONCRETE, for example, includes 200 independent latent units. The presence of a well-trained RBM in the latent space of DVAE is critical to achieve the results quoted in  Table~\ref{tab:1}. In particular, our implementation of DVAE is able to match the result obtained with the CONCRETE method by using only 64+64 latent units rather than 200. A direct demonstration of the necessity to have a well-trained RBM to achieve state-of-the-art performance with DVAE is also given in Table 2 of Ref.~\cite{khoshaman2018gumbolt}.

\begin{table}[t]
\begin{center}
\begin{tabular}{|llll|}
\hline
\multicolumn{4}{|c|}{MNIST (static binarization) }\\
&  &  {\bf ELBO} & {\bf LL} \\
 \hline
DVAE & RBM$_{32\times 32}$ & $ -99.3\pm 0.2$ & $- 90.8\pm 0.2$    \\
  & RBM$_{64\times 64}$ &  $-92.4$ & $ -85.5$  \\
 & RBM$_{128\times 128}$&  $-90.4$ &  $ -84.7$\\
 & RBM$_{256\times 256}$ &  $ \mathbf{-89.2}$ & $ \mathbf{-83.5}$  \\
VIMCO~\cite{mnih2016variational}  & &   &  $ {-91.9}$  \\
 NVIL~\cite{mnih2014neural}  &   &  & $ {-93.5}$  \\
 CONCRETE~\cite{maddison2016concrete} &   &  & $ {-85.7}$  \\
 GS~\cite{jang2016categorical} &   &  ${-101.5}$ &  \\
 RWS~\cite{bornschein2014reweighted}&  & &  ${-88.9}$   \\
 REBAR~\cite{tucker2017rebar} &   &  ${-98.8}$ &  \\
\hline
\end{tabular}
\end{center}
\caption{Comparison of variational generative models with stochastic discrete variables on the validation set of the MNIST dataset. The best results are denoted by boldface font. GS stands for Gumbel-Softmax. The confidence level for the DVAE results is smaller than $\pm0.2$ in all cases.}
\label{tab:1}
\end{table}%


\begin{figure}[t]
\begin{center}
\subfigure[\, Generated MNIST: QVAE, QBM$_{16\times 16}$, $\Gamma = 0$, ]{\includegraphics[width=0.49\columnwidth]{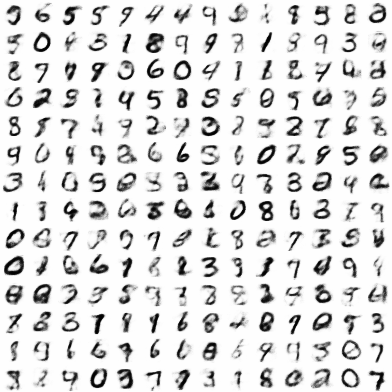}\label{fig:2a}}
\subfigure[\, Generated MNIST: QVAE, QBM$_{64\times 64}$, $\Gamma = 0$, ]{\includegraphics[width=0.49\columnwidth]{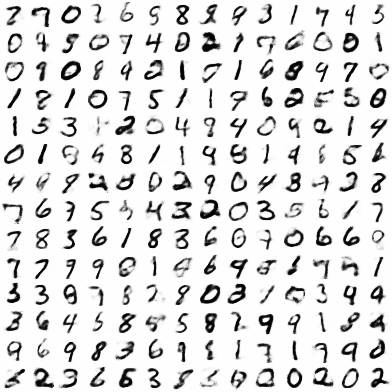}\label{fig:2b}}\\
\subfigure[\, Generated MNIST: QVAE, QBM$_{16\times 16}$, $\Gamma = 1$, ]{\includegraphics[width=0.49\columnwidth]{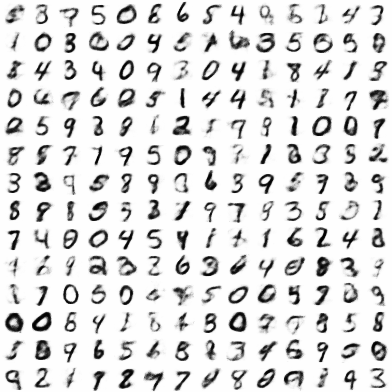}\label{fig:2c}}
\subfigure[\, Generated MNIST: QVAE, QBM$_{64\times 64}$, $\Gamma = 1$, ]{\includegraphics[width=0.49\columnwidth]{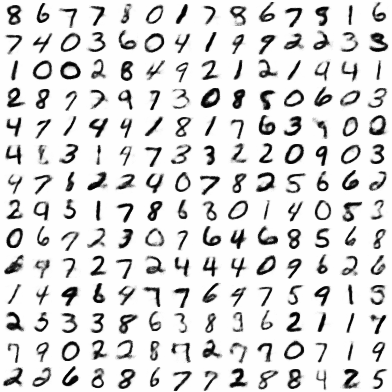}\label{fig:2d}}\\
\subfigure[\, Generated MNIST: QVAE, QBM$_{16\times 16}$, $\Gamma = 2$, ]{\includegraphics[width=0.49\columnwidth]{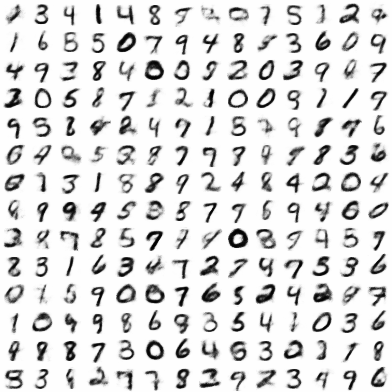}\label{fig:2e}} 
\subfigure[\, Generated MNIST: QVAE, QBM$_{64\times 64}$, $\Gamma = 2$, ]{\includegraphics[width=0.49\columnwidth]{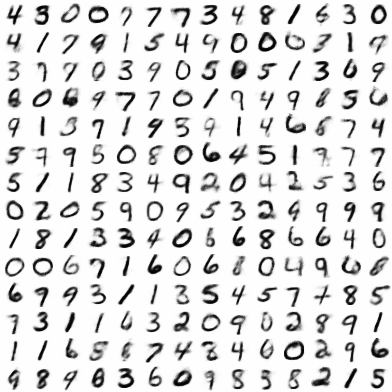}\label{fig:2f}}
\caption{ Comparison of generated MNIST digits with different values of the transverse field $\Gamma$ at the same stage of training (90 epochs).}
\label{fig:2}
\end{center}
\end{figure}

\section{Quantum Variational autoencoders}
\label{sec:QVAE}

We now introduce the QVAE by implementing the prior distribution in the latent space of a VAE  as a QBM. Similar to a classical BM, a QBM is an energy model defined as follows~\cite{Amin:2016qd}:
\bea \label{eq:QBD}
p_\btheta(\z) & \equiv & \tr[\Lambda_\z {e^{-\Ham_\btheta}}]/Z_\btheta\,, \quad Z_\btheta \equiv  \tr [e^{-\Ham_\btheta}]   \,,   \nn
\Ham_\btheta& =&  \sum_l \sigma^x_l \Gamma_l + \sum_l \sigma^z_l h_l + \nn
&  & +  \sum_{l< m}  W_{lm} \sigma^z_l \sigma^z_m, \quad {\bf \Gamma}, {\bf h}, {\bf W}  \in \{\btheta \}\,,
\eea
where $\Lambda_\z \equiv | \z \rangle \langle \z |$ is the projector on the classical state $\z$ and $\sigma_l^{x,z}$ are Pauli operators. States $\z$ are distributed according to $p_\btheta(\z)$; \emph{e.g.}, a quantum Boltzmann distribution for the quantum system given by $\Ham_\btheta$. Similar to the classical case, the ELBO includes the following cross-entropy term:
\bea \label{eq:q_cross_entropy}
H(q_\bphi, p_\btheta ) =  - \E_{\z \sim q_\bphi}[  \log (\tr[\Lambda_\z {e^{-\Ham_\btheta}}])] + \log Z_\btheta\,. 
\eea
Unfortunately, the gradients of the first term in the equation above are intractable. A QBM can still be trained using a  lower-bound to the cross-entropy that can be obtained via the Golden-Thompson inequality~\cite{Amin:2016qd}:
\beq
\tr[ e^{A} e^{B}] \ge \tr [e^{A + B} ]\, \label{GTI},
\eeq
which holds for any two Hermitian matrices. The equality is satisfied if and only if the two matrices commute. Using this inequality we can write for the cross-entropy:
\bea
H(q_\bphi, p_\btheta ) & \ge  &  \overbrace{- \E_{\z \sim q_\bphi}[ \log( \tr[e^{-\Ham_\btheta + \ln \Lambda_\z} ]) ] + \log Z_\btheta}^{\tilde H(q_\bphi, p_\btheta )}   \nn
&=&  \E_{\brho \sim \mathcal{U}}[ \Ham_\btheta(\z(\brho, \bphi))]{+}\log Z_\btheta, \, 
\label{eq:GT1}
\eea
where in the second line we have used the reparameterization trick and the fact that the contribution to the trace of all states different than $\z$ is infinitely suppressed. In the equation above we have defined $\Ham_\btheta(\z) \equiv \bra{\z}  \Ham_\btheta  \ket{\z}$. Using the lower-bound $ \tilde H(q_\bphi, p_\btheta )$, we obtain a tractable  quantum bound (Q-ELBO) to the true ELBO and the QVAE can be trained by estimating the gradients via sampling from the QBM~\cite{Amin:2016qd}:
\beq 
\pd \tilde H(q_\bphi, p_\btheta )  =  \E_{\brho \sim \mathcal{U}}[\pd  \Ham_\btheta(\z(\brho, \bphi))] -\E_{\z \sim p_\theta}[{\pd \Ham_\btheta(\z)}]\,.
\label{eq:GT2}
\eeq
The use of the Q-ELBO and its gradients precludes the training of the transverse fields $\bf \Gamma$~\cite{Amin:2016qd}, which is treated as a constant (hyper-parameter) throughout the training.

\begin{table}[t]
\begin{center}
\begin{tabular}{|lllll|}
\hline
\multicolumn{5}{|c|}{MNIST (static binarization) }\\
 & &  &  {\bf ELBO} &  {\bf Q-ELBO}\\
 \hline
QVAE: & $\Gamma = 0$  & RBM$_{16\times 16}$ &  $ -109.3\pm 0.2$ &   $-109.3\pm 0.2$\\
     &$\Gamma = 1$ &  QBM$_{16\times 16}$	  &     $ -110.5$ &  $-120.6$  \\
    &$\Gamma = 2$  & 			             &   $ -115.3$ &   $- 135.8$\\
QVAE: &$\Gamma = 0$  & RBM$_{32\times 32}$  &  $ -101.8$ &    $-101.8$\\
 &$\Gamma = 1$  &  QBM$_{32\times 32}$ &  $-103.6$ &    $-117.9$\\
           &$\Gamma = 2$  	&				   &  $ -112.1$ &    $-139.7$\\
QVAE: & $\Gamma = 0$  & RBM$_{64\times 64}$     & $-105.7$    &$-105.7$\\
 & $\Gamma = 1$  &   QBM$_{64\times 64}$    & $-108.7$    & $-133.9$\\
            & $\Gamma = 2$  &                                     & $-120.0$    & $ -165.2$\\
\hline
%
%
\end{tabular}
\end{center}
\caption{Evaluation on validation set: RBM$_{16\times 16}$ at 800 epochs,  RBM$_{32\times 32}$ at 250 epochs,  RBM$_{64\times 64}$ at 50 epochs. The confidence level for the numerical results is smaller than $\pm0.2$ in all cases.}
\label{tab:2}
\end{table}%

\subsection{Experimental results with QVAE}
In this section, we show that QVAEs can be effectively trained  via the looser quantum bound (Q-ELBO). While it is computationally unfeasible to train a QVAE that has a large QBM in the latent space with QMC, sampling from large QBMs is possible with the use of quantum annealing devices. Given the results of the previous and present sections, we thus expect the possibility of using quantum annealing devices to sample from large QBMs in the latent space of QVAEs to achieve competitive results on datasets such as MNIST.

To perform experiments with QVAEs, we have considered exactly the same models used in the case of DVAEs, exchanging the RBMs with QBMs. As explicated by Eqs.~\ref{eq:GT1} and~\ref{eq:GT2}, we train the VAE by maximizing the lower bound Q-ELBO to the true ELBO. To compute the negative phase in Eq.~\ref{eq:GT2}, we have used population annealing (PA) for continuous-time quantum Monte Carlo (CT-QMC) (see Appendix~\ref{sec:ctqmc} for more details). We have considered a population of $1000$ samples and $5$ sweeps per gradient evaluation. Despite being one of the most effective sampling methods we considered, population annealing CT-QMC is still numerically expensive and prevented us from fully training QVAEs with large QBMs. We thus considered (restricted) QBMs with $16$, $32$, and $64$ units per layer and estimated the ELBO obtained with different values of the transverse field $\Gamma$. Table~\ref{tab:2} shows the ELBO and Q-ELBO for different sizes of the QBMs and values of $\Gamma$. To estimate Q-ELBO, we use the classical energy function in the positive phase (with no transverse field) and a quantum partition function (see Appendix~\ref{sec:ctqmc}). The ELBO is calculated using the true quantum probability of a state in the positive phase. The difference between ELBO and Q-ELBO is due to Eq.~\ref{GTI}. We emphasize that the results of Table~\ref{tab:2} correspond to ELBOs obtained at different stages during training ($800$, $250$, and $50$ epochs for the cases with $16$, $32$, and $64$ units per layer, respectively). These numbers simply correspond to the largest number of epochs we were able to train each model during the preparation of this work. Also, note that we have chosen not report LL in this table, since this requires using importance sampling which is computationally expensive; this requires calculating the quantum probabilities for a large number of latent samples per input image.

Our results show, as expected, that Q-ELBO becomes looser as the transverse field is increased. Still, we observe that the corresponding ELBO we obtained during training is much tighter and closer to the classical case. This explain why we are able to effectively train QVAE with values of the transverse field as large as 2. We consider this value of the transverse field to be relatively large, since the typical scale of the trained couplings in the classical part of the Hamiltonian is of order 1. Figure~\ref{fig:2} shows MNIST images generated with QVAE and CT-QMC sampling. We see that the quality of the generated samples is satisfactory for the considered values of the transverse field (up to $\Gamma=2$). 

It is important to stress that the deterioration of performance we observe in Table~\ref{tab:2} is mostly due to that fact that the Q-ELBO used for training becomes looser as the transverse field increases. This is not necessarily an intrinsic limitation of QVAE. Indeed, in Ref.~\cite{Amin:2016qd} it was shown that small quantum Boltzmann machines perform better than their classical counterparts if training is performed via direct maximization of the LL.

\section{Conclusions}
\label{sec:conc}
We proposed a variational inference model, QVAE, that uses QBMs to implement the generative process in its latent space. We showed that this infrastructure can be powerful and at higher dimensions of latent space can give state-of-the-art results (for variational inference-based models with stochastic discrete units) on MNIST dataset. We used CT-QMC to sample from the QBMs in the latent space and were limited to smaller dimensions (up to a $64\times 64$ dimensional QBM) due to computational cost. Introduction of QBMs in the latent space of our model introduces an additional quantum bound on the ELBO objective function of VAEs. However, we demonstrated empirically that QVAEs have generally similar performance to their classical limit where the transverse field is absent. On important open question for future work is whether it is possible to improve  the performance of QVAE by using bounds to the LL that tighter than the Q-ELBO used in this work.

During training, both RBM and QBM develop well-defined modes that make sampling via Markov Chain Monte Carlo methods very inefficient. Quantum annealers could provide a computational advantage by exploiting quantum tunnelling to accelerate mixing between different modes. A computational advantage of this type was observed, with respect to quantum Monte Carlo methods, in Ref.~\cite{andriyash2017can}. The successful use of quantum annealers will likely require tailored  implementations that mitigate physical limitations of actual devices such as control errors, limited coupling range and connectivity. This is a promising line of research which we are exploring for upcoming works.

This work is an attempt to add quantum algorithms to powerful existing classical frameworks to obtain new competitive generative models. It can be considered as a bedrock on which the next generation of quantum annealers can be efficiently used in order to solve realistic problems in machine learning. 

\subsection*{Acknowledgements}
The authors would like to thank Hossein Sadeghi, Arash Vahdat, and Jason T. Rolfe for useful discussions during the preparation of this work.

\bibliography{refs}
\appendix
\label{sec:APP}
\section{VAE with Guassian variables}
\label{sec:GaussVAE}

In its simplest version, a VAE's prior and approximate posterior are a product of normal distributions (both with diagonal covariance matrix) chosen as follows:
\bea
p_\btheta(\bzeta) & = &   \mathcal N(\bzeta;{\bf 0}, {\bf 1}) \equiv \prod_{l=1}^L \mathcal N(\zeta_l;0,1)\, \nonumber \\
q_{\bphi}(\bzeta|\x)& = &  \mathcal N(\bzeta; \bm{\mu},\bm{\sigma}) \equiv  \prod_{l=1}^L\mathcal N(\zeta_l;\mu_l,\sigma_l)\,,\nonumber
\eea
where the prior is independent of the parameters $\btheta$ and the means and variances $\bm\mu$ and $\bm\sigma^{2}$ are functions of the inputs $\x$ and of the parameters $\bphi$; the dependence on $\bphi$ is sometimes left implicit when the variable indices are shown. The mean and variance are usually the outputs of a deep neural network. The diagonal Guassians allow for an easy implementation of the reparameterization trick:
\beq
\rho_l \sim \mathcal N(\zeta_l;0,1), \,\,  \zeta_l =\mu_l + \sigma_l \rho \,\,\, \Rightarrow \,\,\, \zeta_l \sim \mathcal N(\zeta_l;\mu_l,\sigma_l)\,. \nonumber
\eeq
The KL divergence is the sum of two simple Guassian integrals $D_{KL}( q_{\bphi}(\bzeta|\x) ||  p_{\btheta}(\bzeta))   =    - H(q_\bphi)  + H(q_\bphi, p_\btheta)$:
\bea
H(q_\bphi)  &  = & - \int q_\bphi(\bzeta) \log q_\bphi(\bzeta) d\bzeta = \ph\ph\nn
 &= &   \frac{1}{2} \sum_{l=1}^L (\log(2\pi) + 1 + \log(\sigma_l^2) ) \nn
 H(q_\bphi, p_\btheta)  &  = & - \int q_\bphi(\bzeta) \log p_\btheta(\bzeta) d\bzeta = \ph\ph\nn
 &= & -  \frac{1}{2} \sum_{l=1}^L (\log(2\pi) + \mu_l^{2} + \sigma_l^{2})\,.
\eea
The only term that requires the reparamaterization trick to obtain a low-variance estimate of the gradient is then the autoencoding term:
\beq
\E_{q_{\bphi}(\bzeta|\x)}[\log p_{\btheta}(\x|\bzeta)] \equiv \E_{\bepsilon}[\log p_{\btheta}(\x|\bm\mu + \bm\sigma\bm\rho)]\,.
\eeq

\section{DVAE with Bernoulli variables}
\label{sec:BERN}
The simplest DVAE can be implemented by assuming that the prior and the approximating posterior are both products of Bernoulli  distributions
\bea
p_\btheta(z_l=1)  & = & p_l \nonumber \\
q_{\bphi}(z_l=1|\x)& = & q_{l}\,,\nonumber
\eea
where the Bernoulli probabilities $q_{l}$ are functions of the inputs $\x$ and of the parameters $\bphi$ and are the outputs of a deep feed-forward network. We have already presented the following expression for the entropy in Sec.~\ref{sec:dvae_repar}:
\bea
H(q_\bphi) \equiv  -  \E_{z \sim q_\bphi}[\log q_\bphi] = \ph\ph\ph\nn
 =   - \sum_{l=1}^L\left(  q_l \log q_l + (1-q_l) \log (1-q_l)  \right)\,.
 \eea
The cross-entropy can be derived similarly:
\bea
H(q_\bphi, p_\btheta)\equiv  -  \E_{z \sim q_\bphi}[\log p_\btheta] = \ph\ph\ph\nn
 =   - \sum_{l=1}^L\left(  q_l \log p_l + (1-q_l) \log (1-p_l)  \right)\,.\,\,\,
 \eea
Similar to the fully Guassian case of the previous section, the only term that requires the reparameterization trick to obtain a low-variance estimate of the gradient is the autoencoding term as in Eq.~\ref{eq:trick}  of the main text.

\section{Hierarchical approximating posterior}
\label{sec:HIER}
Explain-away effects~\cite{goodfellow2016deep} introduce complicated dependencies in the approximating posterior $q_{\bphi}(\bzeta|\x)$, which cannot be fully captured by products of independent distributions as we have considered so far. More powerful variational approximations of the posterior can be considered by including hierarchical structures. In the case of DVAEs, a hierarchical approximating posterior may be chosen as follows:
\beq
q_\bphi(z_l,\zeta_l| \x) =  r(\zeta_l | z_l) q_\bphi(z_l|\zeta_{m<l},\x)\,.
\eeq
A multivariate generalization of the reparameterization trick can be introduced by considering the conditional-marginal CDF defined as follows:
\beq
{\rm F}_l(\zeta_{m\le l}) = \int_{0}^{\zeta_l} q_{\bphi}(\zeta'_l|\zeta_{m<l},\x)d \zeta'_l\,,
\eeq
 where in the expression above we assume the $\zeta_{m\neq l}$ are kept fixed. Thanks to the  hierarchical structure of the approximating posterior, the ${\rm F}_l(\zeta_{m\le l})$ functions are formally the same functions of  $\zeta_l$ and $q_l$ as in the case without hierarchies. The dependence of the functions  ${\rm F}_l(\zeta_{m\le l})$ on the continuous variables $\zeta_{m < l}$ is encoded in the functions  $q_\bphi(\z,\bzeta| \x)$:
 \beq
 {\rm F}_l(\zeta_{m\le l}) =  {\rm F}_l(\zeta_ l, q_l(\zeta_{m <  l}))\,.
 \eeq
 The reparameterization trick is again applied thanks to:
\beq
\rho_l \sim {\mathcal U}, \,\,  \zeta_l =     {\rm F}^{-1}_l(\rho_{m\le l}) \quad \Rightarrow \quad \zeta_l \sim q_{\bphi}(\zeta_l|\x) \nonumber\,.
 \eeq
The KL divergence is:
\bea
&& D_{KL}( q_{\bphi}(\z, \bzeta|\x) ||  p_{\btheta}(\z, \bzeta)) = \nn
& &\ph =  \sum_{l=1}^LD_{KL}(q_\bphi(z_l|\zeta_{m<l},\x) ) ||  p_{\btheta}(z_l)) = \ph\ph\nn
&& \ph =  \sum_{l=1}^L \E_{q_\bphi }[  z_l\log q_l +  (1- z_l) \log (1- q_l)] + \nn
&& \ph - \E_{q_\bphi }[\log p_{\btheta}(\z)]\,. \nonumber
\eea
Notice that, due to the hierarchical structure of the approximating posterior, the expectations above cannot be performed analytically and must be statistically estimated with the use of the reparameterization trick.  

\section{Computing the derivatives}
\label{sec:dervs}
As shown in the previous section, the KL divergence generally includes a term that depends explicitly on the discrete variables $z_l$. When computing the gradients for back-propagation, we must account for the dependence of the discrete variables on the $\bphi$ parameters through the various hierarchical terms of the approximating posterior. Remembering that $z_l =  \Theta(\rho_l+q_l-1)$ and using the chain rule, we have:
\beq
\pd z_l = \pd_{q_{l}}z_l \pd_\bphi q_l =  \delta (\rho_l+q_l-1) \pd q_l\,.
\eeq
The gradient of the expectation over $\rho$ of a generic function of $z$ can then be calculated as follows:
\bea
\pd_\bphi\E_\brho[f(\z)] & = &  \E_\brho[\pd_\bphi f(\z)] = \sum_{l=1}^L \E_{\rho \sim \mathcal{U}}[\pd_{z_l} f(\z) \pd_{q_{l}}z_l \pd_\bphi q_l] = \nn
& = & \sum_{l=1}^L\E_{\rho_{k\neq l}}[\pd_{z_l} f(\z) \pd_\bphi q_l]_{\rho_l = 1-q_l\, \Rightarrow\, z_l = 0} =\nonumber  \\
& = &  \sum_{l=1}^L\E_{\brho}[\pd_{z_l} f(\z)\frac{1-z_l}{1-q_l}\pd_\bphi q_l] = \nn
& = &\sum_{l=1}^L \E_{\brho}[\pd_{z_l} f(\z)(z_l-1)\pd_\bphi \log (1-q_l)]\,,\nonumber
\eea
where, to go from the second to the third row, we have reinstated the expectation over $\rho_l$ by noticing that $q_l$ does not depend on $\rho_l$ and that the condition  $z_l = 0$ may be automatically enforced with the factor $1-z_l$. The term $1-q_l$ accounts for the fact that $z_l=0$ with probability $1-q_l$. This term is necessary to account for the statistical dependence of $z_l$, and thus of $f(\z)$, on variables $z_{m<l}$ that come before in the hierarchy. The equation derived above is useful to compute the derivatives of the positive phase in the case of a hierarchical posterior and an RBM or a QRBM as priors:
\beq
\pd \E_{\brho }\left[ \log p_{\btheta}(\z) \right]  = \E_{\brho}[\pd E_\btheta(\z)] - \E_{p_\theta}[{\pd E_\btheta(\z)}]\,,
\eeq
with
\beq
 f(\z)  = E_\btheta(\z) \quad {\rm or} \quad  \Ham_\btheta(\z)\,.
\eeq

\section{Population-annealed continuous-time quantum Monte Carlo} \label{sec:ctqmc}

To sample from the quantum distribution Eq.~\ref{eq:QBD} we use a continuous-time quantum Monte Carlo  algorithm~\cite{rieger1999application} together 
with a population annealing  sampling heuristic~\cite{hukushima2003population, machta2010population}.

The CT-QMC algorithm is based on the representation of the quantum system with the Hamiltonian Eq.~\ref{eq:QBD} in terms of a classical system with an additional dimension of size $M$ called \emph{imaginary time}.
The classical configuration ${\bf z}$ is replaced with $M$ configurations ${\bf z}^a, a=1,\dots,M$ that are coupled to each other in a periodic manner.
The quantum partition function $Z_\btheta =  \tr [e^{-\Ham_\btheta}] $ can be written as:
 \beq \label{eq:pathInt}
 Z_\btheta \simeq   \sum_{\{ {\bf z}^a\}} \exp \left\{  \log \frac{\Gamma}{M} \sum_{i,a}\frac{1 - z_i^a z_i^{a+1}}{2} -  \frac{1}{M}\sum_{a=1}^M \Ham_0({\bf z}^{a}) \right\}. 
 \eeq
Here $\Ham_0$ is the classical energy $H_0({\bf z}) =    \sum_l z_l h_l +  \sum_{l< m}  W_{lm} z_l z_m$ and periodicity along imaginary time implies ${\bf z}^{M+1}\equiv {\bf z}^1$.

CT-QMC defines a Metropolis-type transition operator acting on extended configurations $T_\btheta:  {\bf z}^a \to {\bf z}^a{'}$. We use cluster updates~\cite{rieger1999application} where clusters
may grow only along the {imaginary time} direction. These updates satisfy detailed balance conditions for the distribution 
 \bea \label{eq:q_distr}
p_\btheta ({\bf z}^a) &=&  e^{-E_q({\bf z}^a) -E_{cl}({\bf z}^a)   }/ Z_\btheta \nonumber\\
E_q({\bf z}^a) &=& -\log \frac{\Gamma}{M} \sum_{i,a}\frac{1 - z_i^a z_i^{a+1}}{2} \nonumber\\
E_{cl}({\bf z}^a)  &=&  \frac{1}{M}\sum_{a=1}^M H_0({\bf z}^{a}).
 \eea
Equilibrium samples from Eq.~\ref{eq:q_distr} allow us to compute the gradient of the bound on the log-likelihood in Eq.~\ref{eq:GT2} as
\bea \label{eq:q_exp}
\E_{p_\theta(\z)}[{\pd \Ham_\btheta(\z)}] = \E_{p_\theta(\z^a)}[{\pd \Ham_\btheta(\z^1)}]\,.
 \eea
To obtain approximate samples from Eq.~\ref{eq:q_distr}, we use PA, which also gives an estimate of the quantum  partition function~\cite{hukushima2003population, machta2010population}. We choose a linear schedule in the space of parameters  
$\btheta_t = t \btheta, t \in [0, 1]$ and anneal an ensemble of $N$ particles $\z^a_{n}, n=1,\dots,N$ with periodic resampling.

Finally, we must evaluate the quantum  cross-entropy Eq.~\ref{eq:q_cross_entropy}, which involves computing probabilities of classical configuration $\bar \z$ under the quantum distribution 
$p_\btheta(\bar \z) \equiv \tr[\Lambda_{\bar \z} {e^{-\Ham_\btheta}}]$. This is done by noticing that 
\bea \label{eq:q_prob}
&& \tr[\Lambda_{\bar \z} {e^{-\Ham_\btheta}}] =  \langle \bar \z |  {e^{-\Ham_\btheta}}  |  \bar \z \rangle \simeq   \nonumber\\
&& \simeq  \sum_{\{ {\bf z}^a, a=2..M\}} \exp \left\{ -E_q({\bf z}^a) -E_{cl}({\bf z}^a) -E_{\rm boundary}({\bf z}^a)  \right\},  \nonumber\\
&& E_{\rm boundary}({\bf z}^a)  =  -\log \frac{\Gamma}{M} \sum_{i}\frac{2 - \bar z_i z_i^{2} - \bar z_i z_i^{M}   }{2}.
 \eea
Thus, to obtain $p_\btheta(\bar \z)$, we must compute the partition function $\langle \bar \z |  {e^{-\Ham_\btheta}}  |  \bar \z \rangle$ of a ``clamped" system, where the first slice of imaginary
 time is fixed $\z^1 \equiv \bar \z$ and we integrate out the rest of the slices taking into account the external field acting on slices $2$ and $M$.

\end{document}